\documentclass[aps,pra,twocolumn,superscriptaddress,floatfix,
nofootinbib,showpacs,longbibliography]{revtex4-1}
\usepackage[utf8]{inputenc}  
\usepackage[T1]{fontenc}     
\usepackage[british]{babel}  
\usepackage[sc,osf]{mathpazo}
\usepackage[scaled=0.86]{berasans}  
\usepackage[colorlinks=true, citecolor=blue, urlcolor=blue]{hyperref}  
\usepackage{graphicx}
\usepackage[babel]{microtype}  
\usepackage{amsmath,amssymb,amsthm,bm,amsfonts,mathrsfs,bbm} 

\usepackage{xspace}  
\usepackage{pgf,tikz}
\usepackage{xcolor}
\usepackage{multirow}
\usepackage{array}
\usepackage{bigstrut}
\usepackage{braket}
\usepackage{color}
\usepackage{natbib}
\usepackage{multirow}
\usepackage{mathtools}
\usepackage{float}
\usepackage{xcolor,colortbl}
\usepackage{physics}
\usepackage{amsmath}
\usepackage{color}
\usepackage[justification=justified, format=plain]{subcaption}
\usepackage[justification=raggedright]{caption}

\newcommand{\be}{\begin{equation}}
\newcommand{\ee}{\end{equation}}
\newcommand{\ba}{\begin{eqnarray}}
\newcommand{\ea}{\end{eqnarray}}

\newtheorem{theorem}{Theorem}

\def\>{\rangle}
\def\<{\langle}

\begin{document}

\title {Genuine entanglement detection via \emph{projection} map in multipartite systems}
	
\author{Bivas Mallick} 
\email{bivasqic@gmail.com}
\affiliation{S. N. Bose National Centre For Basic Sciences, Block JD, Sector III, Salt Lake, Kolkata 700106, India}

\author{Sumit Nandi}
\email{sumit.enandi@gmail.com}
\affiliation{Purandarpur High School, Purandarpur, West Bengal 731129, India}

	\begin{abstract}
We present a formalism to detect genuine multipartite entanglement by considering \emph{projection } map which is a positive but not completely positive map. \emph{Projection } map has been motivated by the “no-pancake theorem” which repudiates the existence of a quantum operation that maps the Bloch sphere onto a disk along its equator. 
The not-complete positivity feature of \emph{projection } map is explored to investigate genuine multipartite entanglement in arbitrary $N$-qubit quantum systems. Our proposed framework can detect some important classes of genuinely entangled states in tripartite and quadripartite scenarios. We provide illustrative example to show the efficacy of our formalism to detect a class of tripartite  PPT bound entangled states. Finally, we construct a suitable witness operator based on \emph{projection } map to certify genuine tripartite entanglement, which is likely to be feasible experimentally.
		
	\end{abstract}
	\maketitle
\section{Introduction}
The enigmatic phenomenon of entanglement, originating from the famous EPR paradox, paves a wonderful way for performing various information processing protocols, and thus, calls for considerable attention for enriching the underlying theoretical foundation. The most fundamental and challenging topic in this direction is to compose a well-defined framework for the detection of entanglement in multipartite systems which provides advantages in several quantum information processing protocols\cite{cleve1999share,raussendorf2001one,
dun2003security,durt2004security}.

To determine whether a given state is entangled has remained a central topic for research. A necessary and sufficient separability criterion for certifying bipartite entanglement is based on positive but not completely positive map. Any bipartite two-qubit state which retains its positivity after the application of partial transposition is known to be separable whereas the one that fails to be a valid density matrix is entangled. 
Thus, positive partial transposition (PPT) criteria is necessary and sufficient for the detection of bipartite entanglement in $2\otimes 2$ and $2\otimes 3$ system \cite{horodecki1996separability}. 
However, multipartite entanglement is much more different notionally as well as operationally from bipartite entanglement \cite{guhne2009entanglement,bhattacharya2021generating}. In this case, there are, in principle, infinite number of ways to partition the constituents into non-overlapping groups. Naturally, multipartite entanglement
is less canonical than its bipartite counterpart. The ubiquitous feature of multipartite entanglement allows infinitely many different stochastic local operations and classical communication (SLOCC) classes\cite{verstraete2002four,de2013maximally} that elucidate the vast structures of multipartite entangled states.  Moreover, in contrast to bipartite cases, many conceptual hinges appear, and the concept of genuine multipartite entanglement (GME) becomes more counterintuitive. \emph{For example}, it has been shown in \cite{bennett1999unextendible} that a state is separable in all of its bipartite cut, yet, not fully separable. In contrast, there exists multipartite states which are entangled across all its bipartitions, nevertheless, the state is not genuinely multipartite entangled \cite{liang2014anonymous,horodecki1995violating}. Thus, it has remained an intimidating task to determine genuineness of entanglement present in a given multipartite state.

Since, detection of multipartite entanglement is a central topic in quantum information theory, several interesting research in this direction has been carried out \cite{guhne2009entanglement,horodecki2009quantum,walter2016multipartite,vaishy2022detecting,clivaz2017genuine,luo2023detecting,xu2023general,li2017measure,hong2021detection,zwerger2019device,baccari2017efficient,zhou2019detecting}. In \cite{horodecki2001separability}, the authors formulated a positive map-based entanglement detection criterion by constructing a $N$-qubit map which is positive on all product states of $N-1$ particles and presented necessary and sufficient conditions for separability of mixed states of $N$-partite states. From different aspects, entanglement detection based on realignment
criterion \cite{rudolph2005further,chen2002matrix} has been proven to be very effective in detecting bipartite entanglement. A suitable generalization of the realignment criterion was proposed in \cite{chen2002generalized} to probe the inseparability of multipartite states. Bloch representation of a $N$-partite state has been found to be a useful scheme to detect multipartite entanglement \cite{hassan2007separability}. Another approach for detecting multipartite entanglement is to construct witness operators \cite{guhne2003investigating,
bourennane2004experimental,toth2005detecting,chruscinski2014entanglement}  which yield a positive expectation value for all biseparable states, and a negative expectation value indicates the presence of genuine entanglement. All these techniques require estimation of a large number of parameters growing rapidly with the number of qubits making genuine multipartite entanglement detection difficult even for a meager system. 

  As an outgrowth of the earlier works \cite{clivaz2017genuine,huber2014witnessing}, here, we propose a framework that provides a necessary criterion for biseparability in multipartite scenario. Our proposed framework is another exposition of not completely positive map-based entanglement detection criterion.  Note that the straightforward extension of not completely positive map operation is not sufficient to detect genuine multipartite entanglement (GME), since, it is only capable of detecting entanglement across its different bipartitions. In \cite{clivaz2017genuine,huber2014witnessing}, the authors circumvent this difficulty by developing a tool that considers convex combinations of not completely positive maps for the detection of GME. Such construction reveals positivity on all biseparable states, but not positivity on the set of genuine entangled states. The authors essentially used transposition map as well as several other well-known maps such as the reduction and the Breuer-Hall map to develop the framework presented therein. 

  In this paper, we consider \emph{projection} map \cite{sudha2021canonical} to detect genuine multipartite entanglement. This map stems from the ``no-pancake theorem" \cite{blume2010information} which states that there does not exist a quantum operation that kills one component of polarization, thereby mapping the entire Bloch sphere onto a disk touching the sphere. By virtue of it, \emph{projection} map reduces the number of parameters required to characterize a state on the Bloch sphere. It motivates us to construct a general framework for certifying GME by taking convex combination of the \emph{projection} map which is essentially a positive but not completely positive map. We further demonstrate that by selecting appropriate parametrization, the \emph{projection} map can be generated from time independent Lindblad operators. We discover the novelty of \emph{projection} map in detecting entanglement, specifically genuine multipartite entanglement which has not been discussed earlier. It has been shown that \emph{projection} map can detect both inequivalent SLOCC classes of tripartite genuine entanglement. Further, we have shown that our proposed map has the potential to detect a particular class of tripartite bound entangled state and a generic SLOCC class of quadripartite state. To further explore the operational advantages of the \emph{projection} map, we construct a tripartite entanglement witness and provide its decomposition into experimentally realizable local observables.
  
The plan of the paper is organized as follows. In Sec.\ref{secII}, we discuss about \emph{projection} map, its positivity and not complete positivity. We further show that this map arises from the Lindblad structure. Then we present a framework for genuine multipartite entanglement detection using \emph{projection} map in Sec.\ref{secIII}. In Sec.\ref{secIV}, we demonstrate the effectiveness of \emph{projection} map for detecting genuine multipartite entanglement by considering several examples in tripartite and quadripartite scenarios. Finally, we conclude in (Sec.\ref{conclusion}) along with some interesting future perspectives.
  \section{\emph{Projection} map in $M_2$} \label{secII}
 Let $\mathbb{C}^d$ be the complex Hilbert space of dimension $d$ and  $B(\mathbb{C}^d)$ be the set of all bounded operators acting on $\mathbb{C}^d$.   The operators acting on a finite-dimensional Hilbert space can be represented as matrices. We define the set of $d \times d$ complex matrices by $M_d$. The state space of a qubit is often visualized by the Bloch sphere representation in $\mathbb{C}^2$ as follows
  \begin{equation}
\rho = \frac{1}{2}\Big(I_2+\sum_{i=1}^3 p_i\sigma_i\Big)\\ 
 \end{equation}

where $I_2$ is the $2\times2$ identity matrix and $\sigma_1$, $\sigma_2$, $\sigma_3$ are pauli matrices written explicitly as 
\begin{equation} \label{pauli}
  \sigma_1= \begin{bmatrix} 
	0& 1\\  
	1& 0 \\
	\end{bmatrix}, \sigma_2= \begin{bmatrix} 
	0& -i \\
	i& 0 \\
	\end{bmatrix},  \sigma_3= \begin{bmatrix} 
	1& 0 \\
	0& -1 \\ 
	\end{bmatrix} 
\end{equation}
The component of Bloch vectors $p_i$ have the following properties: $p_i\in\mathbb{R}^3$, $p_i=Tr(\rho\sigma_i)$ and $\sum_{i=1}^3 |p_i|^2\le 1$.  As mentioned earlier, the \emph{projection} map arises from the generalization of ``no-pancake theorem" which states that the image of the Bloch sphere cannot be a disk that touches the sphere \cite{blume2010information}. \emph{Projection} map annihilates $\sigma_3$ component while preserving the other Bloch components. Therefore the action of map $\mathbb{P}$ can be written as \cite{sudha2021canonical}: 
\begin{equation}
    \mathbb{P}(p_1,p_2,p_3)^T \rightarrow (p_1,p_2,0)^T
\end{equation}
We write the qubit density matrix $\rho$ explicitly in the standard basis 
$\ket{0}=(1, 0)^T$ and $\ket{1}=(0, 1)^T$ as follows
\begin{equation}
    	\rho = \begin{bmatrix} 
	\frac{1+p_3}{2} &\frac{p_1-ip_2}{2} \\[0.2cm]
 \frac{p_1+ip_2}{2}& \frac{1-p_3}{2}\\ 
	\end{bmatrix}
\end{equation} 
The \emph{projection} map is defined as 
\begin{equation}\label{aftermap}
    	\mathbb{P}(\rho) = \begin{bmatrix} 
	\frac{1}{2} &\frac{p_1-ip_2}{2} \\[0.2cm]
	\frac{p_1+ip_2}{2}& \frac{1}{2}\\ 
	\end{bmatrix}
	\end{equation}
For arbitrary $X\in M_2$,  
\begin{equation}
  X= \begin{bmatrix} 
	x_{11}& x_{12}\\  
	x_{21}& x_{22} \\
		\end{bmatrix}\nonumber, \{x_{ij}\}\in \mathbb{C}
	\end{equation}
The action of  \emph{projection} map $\mathbb{P}: M_2\rightarrow M_2$ can be written as:
 \begin{equation}\label{proj}
  \mathbb{P}(X)= \begin{bmatrix} 
\frac{x_{11}+x_{22}}{2}& x_{12} \\[0.2cm] 
	x_{21}& \frac{x_{11}+x_{22}}{2} \\
		\end{bmatrix}
			\end{equation}
One can easily check from Eq.(\ref{proj}) that the \emph{projection} map is linear.
\subsection{Positivity and not complete positivity of \emph{projection} map}
\textbf{Positivity:}
To verify positivity of the \emph{projection} map $\mathbb{P}: M_2\rightarrow M_2$, we have to show that ${\mathbb{P}}[\rho] \ge 0$ for all $\rho\ge 0$. Therefore, it is sufficient to show that all the principal minors of $\mathbb{P}(\rho)$ in Eq.\eqref{aftermap} are positive. The diagonal elements of $\mathbb{P}(\rho)$ are the first order principal minor which are clearly positive numbers, and the value of second order principal minor is $\frac{1-(|p_1|^2+|p_2|^2)}{4}$. Using the fact $\sum_{i=1}^3 |p_i|^2\le 1$, it can be checked that $\frac{1-(|p_1|^2+|p_2|^2)}{4}\geq 0$, which proves the positivity of our proposed map $\mathbb{P}$. \\ \qed

\textbf{Not complete positivity:}
 To show that \emph{projection} map is not a completely positive map, one can make use of Choi Jamiolkowski isomorphism \cite{jamiolkowski1974effective,choi1975positive}. In the computational basis $\{\ket{i}\}$, the maximally
entangled state is $\ket\phi=\frac{1}{\sqrt 2}(\ket{00}+\ket{11})$. By Choi- Jamiolkowski isomorphism, it is sufficient to show that, the Choi matrix $  C^{\mathbb{P}}=\mathbb{I} \otimes \mathbb{P} (\ket\phi\bra\phi)$ is not positive semidefinite. By computing the corresponding Choi matrix, we get \begin{equation}
	    C^{\mathbb{P}} =
	\begin{bmatrix} 
	\frac{1}{4}& 0 &	0 &	\frac{1}{2}\\[0.1cm]
 0  & 	\frac{1}{4} & 0&0\\[0.1cm]
	0& 0 &	\frac{1}{4} &	0\\[0.1cm]
	\frac{1}{2}& 0 &0&	\frac{1}{4}\\[0.1cm]
	\end{bmatrix}\\
	\end{equation}
We obtain a negative eigenvalue $-\frac{1}{4}$ of the above Choi matrix which implies that $\mathbb{P}$ is not a completely positive map. 
 \subsection{Generating \emph{projection} map from Lindblad structure}
 Here, we generate \emph{projection} map from a well-defined physically realizable structure of open quantum systems. Isolated systems undergo unitary evolution. However, the evolution of a general quantum system is described by completely positive and trace preserving  (CPTP) maps. The generators of these CPTP maps are associated with Lindblad-type super operators \cite{alicki2007quantum,lindblad1976generators,breuer2002theory,rivas2014quantum,rivas2010entanglement,breuer2009measure,bhattacharya2020convex,bhattacharya2017exact,mallick2024assessing}. Importantly, it is possible to construct positive maps from Lindblad operators by choosing suitable parametrization \cite{hall2014canonical}. Here, we show that we can generate \emph{projection} map from the time-independent Lindblad structure. The corresponding map $\Lambda: M_2\rightarrow M_2$ is 
\begin{equation} \label{lmap}
    \Lambda(X)= (\mathbb{I} + \mathcal{L})\hspace{0.05cm}(X) \hspace{0.2cm} \text{for all} \hspace{0.2cm} X =\begin{bmatrix} 
	x_{11}& x_{12}\\  
	x_{21}& x_{22}
\end{bmatrix} \in M_2
\end{equation}
with, 
\begin{equation}
   \mathcal{L} (X) =  \sum_{i} \gamma_i (\sigma_i X {\sigma_i}^{\dagger} - \frac{1}{2}({\sigma_i}^{\dagger}  \sigma_i X + X {\sigma_i}^{\dagger}  \sigma_i ))
\end{equation}
where, $\gamma_i$ are the time independent Lindblad coefficients and $\sigma_i$'s are pauli matrices defined in Eq.\eqref{pauli}.

We take, $\gamma_1=\gamma_2=\frac{1}{4}$  and $\gamma_3=-\frac{1}{4}$, then Eq.\eqref{lmap} becomes 
\begin{equation} \label{lindbladmap}
\begin{split}
    \Lambda(X) &= X + \frac{1}{4} (\sigma_1 X {\sigma_1} - \frac{1}{2}{\sigma_1}  \sigma_1 X -\frac{1}{2} X {\sigma_1} \sigma_1 ) \\ &
   + \frac{1}{4} (\sigma_2 X {\sigma_2} - \frac{1}{2}{\sigma_2}  \sigma_2 X -\frac{1}{2} X {\sigma_2} \sigma_2 ) \\ &
   -  \frac{1}{4} (\sigma_3 X {\sigma_3} - \frac{1}{2}{\sigma_3}  \sigma_3 X -\frac{1}{2} X {\sigma_3} \sigma_3 ) 
    \end{split}
\end{equation}
After simplification, Eq.\eqref{lindbladmap} turns out to be
\begin{equation}\label{aciton}
     \Lambda(X) = \begin{bmatrix} 
	\frac{x_{11}+x_{22}}{2}& x_{12} \\[0.2cm] 
	x_{21}& \frac{x_{11}+x_{22}}{2} \\
		\end{bmatrix}
\end{equation}
Therefore, for specific values of the time independent Lindblad coefficients $\gamma_i$, we can generate \emph{projection} map $(\mathbb{P})$. 
\section{Framework for genuine multipartite entanglement detection using \emph{Projection} map }\label{secIII}
The notion of separability is less unambiguous in a multipartite scenario compared to its bipartite counterpart. For instance, separability may exist in one particular bipartition amongst its numerous inequivalent partitions. Here we consider two operationally distinct frameworks of separability. An $N$-partite state is called fully separable iff it can be decomposed into its constituents
 \begin{equation}
        \rho_{sep}  = \sum_{i=1}^N p_{i} {\rho_{1}}^{i} \otimes ....\otimes {\rho_{N}}^i
    \end{equation}
    where $\{p_i\}$ is a probability distribution satisfying $\sum_{i=1}^N p_i=1$. If a state cannot be written in the above way, then we can say that the system must contain some entanglement. However, here we consider a particular case, as follows: a state $\rho_{2-sep}$ is biseparable iff it can be decomposed as
\begin{equation}\label{bisep}
\rho_{2-sep} = \sum_{A} \sum_{i} {p_{A}}^{i} {\rho_{A}}^{i} \otimes {\rho_{\Bar{A}}}^{i}  
\end{equation}
 where $\rho_A$ denotes a quantum state for the subsystem defined by the subset A and $\sum_{A}$ stands for the sum over all bipartitions $A|\Bar{A}$. An $N$-partite state that can not be decomposed as Eq.(\ref{bisep}) is called a genuine multipartite entangled (GME) state \cite{clivaz2017genuine}. It suggests that entanglement persists across all of its possible $2^{N-1}-1$ inequivalent bipartitions.

In the following subsections, we shall introduce a methodology to detect genuine multipartite entanglement via \emph{projection} map. 
 \subsection{\emph{Projection} map in tripartite system }
 Here, we will present a framework for detecting genuine tripartite entanglement using \emph{projection} map.  In order to achieve this goal, we start with the simplest tripartite scenario by considering the map 
	 \begin{equation}
    \begin{split} \label{tripartitemap}
   \Phi_{3}({\varrho}) &=  [\mathbb{P}_{1} \otimes \mathbb{I}_{2} \otimes \mathbb{I}_{3}+ \mathbb{I}_{1} \otimes \mathbb{P}_{2}  \otimes \mathbb{I}_{3} + \mathbb{I}_{1} \otimes \mathbb{I}_{2} \otimes \mathbb{P}_3\\
     &  + \kappa_3.\hspace{0.1cm} I \hspace{0.1cm} .Tr]({\varrho}) 
   \end{split}
\end{equation}
	  where $\mathbb{P}_{i}$ denotes the action of \emph{projection} map on the $i$-th qubit,              $\rho$ is an arbitrary state in $\mathbb{C}^2\otimes \mathbb{C}^2 \otimes \mathbb{C}^2$ and $\kappa_3$ is a constant. Note that $\kappa_3$ plays a crucial role in determining the state $\rho$. For this reason, $\kappa_3$ has to be chosen such that 
	 \begin{equation} \label{positiveonbisep}
	     \Phi_{3}(\rho_{2-sep}) \geq 0
	 \end{equation}
\emph{iff} $\rho_{2-sep}$ is a biseparable state in   $\mathbb{C}^2\otimes \mathbb{C}^2 \otimes \mathbb{C}^2$  and can be expressed as Eq.(\ref{bisep}). To determine an appropriate value for 
$k_3$, we will first prove the following proposition: 

 \textbf{Proposition 1:} The minimum eigen value of ${\mathbb{I}}_{{A}} \otimes {\mathbb{P}}_{B}$ when acting on any two-qubit state is $-\frac{1}{4}$, which is achieved for a maximally entangled state.
\begin{proof}
To validate this statement, we must first revisit the properties of the minimal output eigenvalues of positive maps \cite{clivaz2017genuine}. Let, ${\eta_{min}}(\Lambda)$ be the minimum eigenvalue  of a positive map $\Lambda$ acting on $M_2$, i.e.
\begin{equation}
    {\eta_{min}}(\Lambda) = \text{min}_{\sigma_{AB}} {EV}_{min} (({\mathbb{I}}_{{A}} \otimes {\Lambda}_{B}) \hspace{0.1cm} \sigma_{AB})
\end{equation}
where, $EV_{min}$ denotes the minimum eigenvalue. Note that, to determine ${\eta_{min}}(\Lambda)$, it is sufficient to minimize only over pure states \cite{clivaz2017genuine}. Thus for \emph{projection} map $({\mathbb{P}})$, $ {\eta_{min}}({\mathbb{P}})$ is obtained by minimizing over two-qubit pure states. Since, \emph{projection} map is a positive map, therefore, for pure product states the minimum eigenvalue is always non-negative. Therefore, to determine $ {\eta_{min}}({\mathbb{P}})$, it is sufficient to find the minimum eigenvalue only over pure entangled states. We can consider such entangled states in the following form:
\begin{equation}
    \ket{\psi} = \sqrt{\nu_1} \ket{00} + \sqrt{\nu_2} \ket{11}  \hspace{0.2cm} \text{where,} \hspace{0.2cm} |\nu_1| + |\nu_2| =1
\end{equation}
By computing the minimum eigenvalue of $({\mathbb{I}}_{{A}} \otimes {\mathbb{P}}_{B}) (\ket{\psi} \bra{\psi})$, we get
\begin{equation} \label{minimumeigenvalue}
     {\eta_{min}}(\mathbb{P}) = \frac{1}{4} \Big[|\nu_1|  + |\nu_2| - \sqrt{{|\nu_1|}^2 + 14 \hspace{0.1cm} |\nu_1| |\nu_2|+ {|\nu_2|}^2} \Big]
\end{equation}
using $|\nu_1| + |\nu_2| =1$, Eq.\eqref{minimumeigenvalue} becomes
\begin{equation}
   {\eta_{min}}(\mathbb{P}) = \frac{1}{4} \Big[ 1 - \sqrt{ 1 + 12 \hspace{0.1cm} {|\nu_1|}{|\nu_2|}} \Big]
\end{equation}\\

Minimum value of $ {\eta_{min}}(\mathbb{P})$ occurs when ${|\nu_1|}{|\nu_2|}$ is maximum, which is obtained when $\sqrt{\nu}_1 = \sqrt{\nu}_2 = \frac{1}{\sqrt{2}}$ i.e. when $\ket{\psi}$ is a maximally entangled state and the minimum eigenvalue reduces to $ - \frac{1}{4}$.
\end{proof}

Using the above proposition, now we will prove the following theorem:
\begin{theorem}
For $\kappa_3 = \frac{1}{2}$, $\Phi_{3}$ is positive on all biseparable states
\begin{equation}
\Phi_{3}(\rho_{2-sep})\geq 0
\end{equation}
 \end{theorem}\label{th1}
  \begin{proof} Following Eq.(\ref{bisep}), we write a biseparable state in the most general form as 
  \begin{equation}\label{3bisep}
  \rho_{2-sep} = p_1 \hspace{0.1cm}\rho_1 \otimes \rho_{23} + p_2 \hspace{0.1cm} \rho_2 \otimes \rho_{13}+p_3 \hspace{0.1cm} \rho_3 \otimes \rho_{12}
  \end{equation}
Now, applying the map $ \Phi_{3}$ on $\rho_{2-sep}$, we get
\begin{equation}
\begin{split}
 & \Phi_{3}(\rho_{2-sep}) \\
 & = (\mathbb{P}_1\otimes \mathbb{I}_{2} \otimes \mathbb{I}_{3})( p_1 \rho_1 \otimes \rho_{23} +
	 p_2 \rho_2 \otimes \rho_{13}+p_3 \rho_3 \otimes \rho_{12}) \\ 
  & + (\mathbb{I}_{1} \otimes \mathbb{P}_2 \otimes \mathbb{I}_{3})( p_1 \rho_1 \otimes \rho_{23} + 
	 p_2 \rho_2 \otimes \rho_{13}+p_3 \rho_3 \otimes \rho_{12}) \\ 
  & + (\mathbb{I}_{1} \otimes \mathbb{I}_{2} \otimes \mathbb{P}_3)( p_1 \rho_1 \otimes \rho_{23} + p_2 \rho_2 \otimes \rho_{13}+p_3 \rho_3 \otimes \rho_{12})  \\
  & +\frac{I}{2} 
 \end{split}
\end{equation}

In order to prove our theorem, we have to show that $\Phi_{3}(\rho_{2-sep})$ is always positive semidefinite. To do so first we evaluate the minimum eigenvalue of $\Phi_{3}(\rho_{2-sep})$. Let us denote $\widetilde{\lambda_{min}}$ be the minimum eigenvalue of ${\Phi_{3}}(\rho_{2-sep})$. Now using Proposition 1, we can write 

\begin{eqnarray}
\begin{split}
& \widetilde{{\lambda_{min}}}({\Phi_{3}}(\rho_{2-sep}))  \nonumber \\
& =\widetilde{\lambda_{min}}\Big((\mathbb{P}_1\otimes \mathbb{I}_{2} \otimes  \mathbb{I}_{3})( p_1 \rho_1 \otimes \rho_{23} +
	 p_2 \rho_2 \otimes \rho_{13}+p_3 \rho_3 \otimes \rho_{12})\nonumber\\
	& \hspace{0.5cm}+ (\mathbb{I}_{1} \otimes \mathbb{P}_2  \otimes \mathbb{I}_{3})( p_1 \rho_1 \otimes \rho_{23} +  p_2 \rho_2 \otimes \rho_{13}+p_3 \rho_3 \otimes \rho_{12})\nonumber\\
	& \hspace{0.5cm} +( \mathbb{I}_{1} \otimes \mathbb{I}_{2} \otimes \mathbb{P}_3)( p_1 \rho_1 \otimes \rho_{23} + p_2 \rho_2 \otimes \rho_{13}+p_3 \rho_3 \otimes \rho_{12}) \nonumber\\
 & \hspace{0.5cm} + \frac{I}{2} \Big) \\
	 & \geq	 \Big((\mathbb{P}_1\otimes \mathbb{I}_{2} \otimes \mathbb{I}_{3})( \underbrace{p_1 \rho_1 \otimes \rho_{23}}_{\geq0} +
	 \underbrace{p_2 \rho_2 \otimes \rho_{13}}_{\geq -\frac{1}{4}}+\underbrace{p_3 \rho_3 \otimes \rho_{12}}_{\geq -\frac{1}{4}}) \nonumber    \\ 
	& \hspace{0.5cm} + (\mathbb{I}_{1} \otimes \mathbb{P}_2  \otimes \mathbb{I}_{3})(\underbrace{ p_1 \rho_1 \otimes \rho_{23}}_{\geq-\frac{1}{4}} + 
	\underbrace{ p_2 \rho_2 \otimes \rho_{13}}_{\geq 0}+\underbrace{p_3 \rho_3 \otimes \rho_{12}}_{\geq -\frac{1}{4}}) \nonumber\\
	& \hspace{0.5cm}+ (\mathbb{I}_{1} \otimes \mathbb{I}_{2}\otimes \mathbb{P}_3 )( \underbrace{p_1 \rho_1 \otimes \rho_{23}}_{\geq -\frac{1}{4}} + \underbrace{p_2 \rho_2 \otimes \rho_{13}}_{\geq -\frac{1}{4}}+\underbrace{p_3 \rho_3 \otimes \rho_{12}}_{\geq 0}) \nonumber   \\
 & \hspace{0.6cm} +\underbrace{\frac{I}{2}}_{\geq \frac{1}{2}}\Big) \\
&\geq 2\sum_{i=1}^3p_i(-\frac{1}{4})+\frac{1}{2}\nonumber\\
&\geq 0\nonumber\\
\end{split}
\end{eqnarray}

In the last line, we have used the fact that $\sum_{i=1}^3p_i=1$. This concludes the proof of our theorem. 
\end{proof}
Note that, this value of $\kappa_{3}$ is optimal. Optimality can be achieved by considering a biseparable state, \emph{for example} $\ket{{\phi}^{+}}\otimes\ket{0}$ where, $\ket{{\phi}^{+}}$ is maximally entangled state in ${\mathbb{C}}^2 \otimes {\mathbb{C}}^2$, and thereby applying the given map $\Phi_{3}$ on it. The corresponding value of $\kappa_3$ turns out to be $\frac{1}{2}$, which is the same as obtained earlier.

\subsection{$N$-partite generalization of \emph{projection} map}
We will now envisage the inseparability of an arbitrary N-partite state by directly lifting $\mathbb{P}$. We denote $A\subset\{1,2,\dots,n\}$ be the proper subset of the parties. For a given N, there are $2^{N-1}-1$ number of inequivalent ways to arrange all bipartitions. Hence lifting of $\mathbb{P}$ involves convex combination of all such terms. Let us define the following map:
\begin{equation}\label{nmap}
{\Phi_{N}} (*)=\left[\sum_A\mathbb{P}_A \otimes
\mathbb{I}_{\bar{A}}+\kappa_N .  {I}.Tr\right](*)
\end{equation}  
\\  	
Now we present the following theorem:	
\begin{theorem}
For $\kappa_N=\frac{1}{4}\times(2^{N-1}-2)$, $\Phi_{N}$ is positive on all N-partite biseparable states $\rho_{2-sep}$ given by Eq.(\ref{bisep}), 
\begin{equation}\label{th2}
\Phi_{N}(\rho_{2-sep})\geq 0
\end{equation}
  \end{theorem}

\textit{proof:} To prove the above theorem, we apply Eq.(\ref{nmap}) on the biseparable state given by Eq.(\ref{bisep}) and evaluate minimum eigenvalue $(\mu_{min})$ of $\Phi_{N}(\rho_{2-sep})$ 
\begin{equation}
    \begin{split}
       & \mu_{min}(\Phi_{N}[\rho_{2-sep}]) \\
       & =   \mu_{min}\Big(\sum_A\mathbb{P}_A \otimes
\mathbb{I}_{\bar{A}}+\kappa_N. I.Tr\Big)[\rho_{2-sep}]  \\ &
=\mu_{min}\Big(\sum_A\mathbb{P}_A \otimes
\mathbb{I}_{\bar{A}}+\kappa_N. I.Tr\Big)\sum_{A^\prime} \sum_{i} {p_{A^\prime}}^{i} {\rho_{A^\prime}}^{i} \otimes {\rho_{\Bar{A^\prime}}}^{i} \\ & 
\ge \sum_{A^\prime}\sum_i\Big( \sum_{A=A^\prime}p_i^{A^\prime}\underbrace{\mu_{min}(\mathbb{P}_{A^\prime}\otimes \mathbb{I}_{\bar{A}^\prime}[\rho_i^{A^\prime}\otimes\rho_i^{\bar{A}^\prime}])}_{\ge 0}\\
& \hspace{1cm}+ \sum_{A\neq A^\prime}p_i^{A^\prime}\underbrace{\mu_{min}(\mathbb{P}_{A^\prime}\otimes \mathbb{I}_{\bar{A}^\prime}[\rho_i^{A^\prime}\otimes\rho_i^{\bar{A}^\prime}]}_{\ge -\frac{1}{4}})\Big) \\
& \hspace{1cm} + (2^{N-1}-2)\times \frac{1}{4}\underbrace{\mu_{min}(I}_{=1})\\
& \ge 0
    \end{split}
\end{equation}
Since the sum $\sum_{A\neq A^\prime}$ involves $(2^{N-1}-2)$ terms, we obtain the inequality in the last line. \qed

Note that we obtain the corresponding tripartite map as given in Eq.(\ref{tripartitemap}) by substituting $N=3$ in our generalized map for the $N$-partite state given by Eq.(\ref{nmap}).
To illustrate the further application of our presented map $\Phi_{N}$, we now consider a particular case for $N=4$.  A quadripartite system involving four subsystems gives rise to seven inequivalent bipartitions occurring across $1|234$, $2|134$, $3|124$, $4|123$, $12|34$, $13|24$, and $14|23$.  

Considering all these different bipartitions of a quadripartite state, we define a map $\Phi_{4}()$ as follows
\begin{equation}\label{quadripartitemap}
    \begin{split}
 \Phi_{4}(\omega) & = \big(\mathbb{P}_{1} \otimes \mathbb{I}_{2} \otimes \mathbb{I}_{3} \otimes \mathbb{I}_{4} + \mathbb{I}_{1} \otimes \mathbb{P}_{2}  \otimes \mathbb{I}_{3} \otimes \mathbb{I}_{4} \\ 
& +\mathbb{I}_{1} \otimes \mathbb{I}_{2} \otimes \mathbb{P}_3 \otimes \mathbb{I}_{4} +   \mathbb{I}_{1} \otimes \mathbb{I}_{2}  \otimes \mathbb{I}_{3} \otimes \mathbb{P}_{4} \\
& + \mathbb{P}_{1} \otimes \mathbb{P}_{2} \otimes \mathbb{I}_{3} \otimes \mathbb{I}_{4} + \mathbb{P}_{1} \otimes \mathbb{I}_{2} \otimes \mathbb{P}_{3} \otimes \mathbb{I}_{4}  \\ 
& + \mathbb{P}_{1} \otimes \mathbb{I}_{2} \otimes \mathbb{I}_{3} \otimes \mathbb{P}_{4} + \kappa_4 .\hspace{0.1cm}  I .\hspace{0.1cm} Tr\big)(\omega). 
\end{split}
\end{equation}
 where $\omega$ is an arbitrary state in $\mathbb{C}^2\otimes \mathbb{C}^2 \otimes \mathbb{C}^2 \otimes \mathbb{C}^2$ and $\kappa_4$ is a constant. We choose $\kappa_4$ in such a way that 
 \begin{equation}
     \Phi_{4}(\rho_{2-sep}) \geq 0
 \end{equation}
iff $\rho$ is biseparable state in $\mathbb{C}^2\otimes \mathbb{C}^2 \otimes \mathbb{C}^2 \otimes \mathbb{C}^2 $ and can be expressed in the form as given in Eq.(\ref{bisep}).\\~\\
\textbf{Lemma1:} For $\kappa_4= \frac{3}{2}$, $\Phi_{4}$ is positive for all 4-partite biseparable states
which can be expressed in the form \eqref{bisep} i.e.
\begin{equation}\label{th1}
\Phi_{4}(\rho_{2-sep})\geq 0
\end{equation}

Proof of the above Lemma follows from the proof of Theorem 2.

In the next section, we will show the effectiveness of the \emph{projection} map for certifying GME states.
\section{Efficacy of \emph{projection} map for the detection of entangled states} \label{secIV}
Before exploring multipartite states, we will first examine the efficacy of the \emph{projection} map in detecting entanglement of a bipartite two-qubit state in the presence of noise.\\

\textbf{Bipartite system:}
Consider, the Werner state given by:
\begin{equation}
\rho_w=p|\phi\rangle\langle\phi|+\frac{1-p}{4}I_{2\otimes 2},
\end{equation}
where the parameter $p \in [0,1]$ indicates the presence of white noise. ${\rho_{w}}$ reduces to maximally mixed separable state for $p=0$ whereas for $p=1$, it denotes maximally entangled state. It is known that ${\rho_{w}}$ features entanglement if $p>\frac{1}{3}$ and it violates Bell-CHSH inequality for $p>\frac{1}{\sqrt2}$ \cite{horodecki1995violating}. Now, the action of \emph{projection} map defined in Eq.\eqref{proj} on ${\rho_{w}}$ produces the following matrix:
\begin{equation}
	 \big(\mathbb{I}\otimes \mathbb{P}\big)  {\rho_{w}}=
	\begin{bmatrix} 
	\frac{1}{4}& 0 &0 &	\frac{p}{2}\\[0.1cm]
    0& 	\frac{1}{4} & 0&0\\[0.1cm]
	0& 0 &	\frac{1}{4} &	0\\[0.1cm]
	\frac{p}{2}& 0&0&	\frac{1}{4}\\[0.1cm]
	\end{bmatrix}\\
	\end{equation}
The four eigenvalues of $\big(\mathbb{I}\otimes \mathbb{P}\big) {\rho_{w}}$ are $\frac{1}{4}$, $\frac{1}{4}$, $\frac{1}{4}(1-2p)$ and $\frac{1}{4}(1+2p)$, respectively. It can be checked easily that for $p>0.5$, the eigenvalue $\frac{1}{4}(1-2p)$ is negative. Thus, the action of \emph{projection} map yields a bound on the visibility parameter $p$ beyond which the Werner state is entangled. In the context of thermodynamical work extraction protocol, the authors \cite{maruyama2005thermodynamical} had presented a novel separability criterion, and shown that ${\rho_{w}}$ is entangled if $p>0.60$. Remarkably, our presented formalism is capable of detecting entanglement in Werner state for a wider range of the visibility parameter.\\

\textbf{Tripartite system:}
We consider GHZ state which is a genuine entangled state 
\begin{equation}
\ket{GHZ}=\frac{1}{\sqrt 2}(\ket{000}+\ket{111}).
\end{equation} 	     	  
Evaluating the corresponding  $\Phi_{3}(\ket{GHZ} \bra{GHZ})$ for GHZ state using the map defined in Eq.(\ref{tripartitemap}), we get a negative eigenvalue $- 0.25$. Note that, previous studies in this direction by taking transposition map into consideration \cite{clivaz2017genuine,chimalgi2023detecting}, were only able to detect GHZ state after performing $\sigma_x$ rotation onto the map. Therefore, \emph{projection} map exempts us of applying an additional unitary operation for detecting GHZ state. \\

Moreover, to investigate the robustness of \emph{projection} map against noise, we consider noisy GHZ state as follows
\begin{equation}
\rho_{noisy}=x|GHZ\rangle\langle GHZ|+\frac{1-x}{8}{I}_8.
\end{equation} 	
The map defined in Eq.\eqref{tripartitemap} i.e.  $\Phi_{3}$ can detect $\rho_{noisy}$. It can be checked that the threshold value of $x$ for certifying GME of the given state turns out to be $x > 0.78$. \\

\begin{figure}[ht]
\includegraphics[width=.35\textwidth]{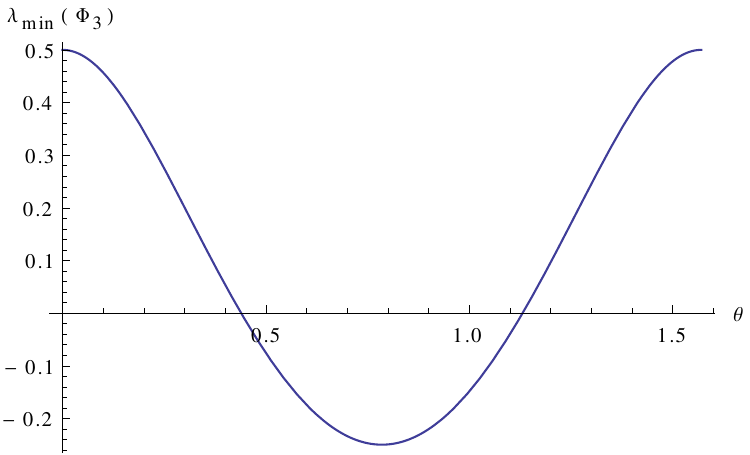}
\caption{ The negative eigenvalue $({\lambda_{min}}(\Phi_3))$ of $\Phi_3(\ket{{GHZ}_g} \bra{{GHZ}_g})$ corresponding to the generalized GHZ state is plotted against the state parameter $\theta$ measured in radians.}\label{gen_ghz}
\centering
\end{figure}

We consider generalized tripartite GHZ state given by $\ket{GHZ}_g=\cos\theta\ket{000}+
\sin\theta\ket{111}$. We show the minimum eigenvalue of the corresponding Choi matrix $\lambda_{min}(\Phi_3)$  as a function of the state parameter $\theta$ in Fig.(\ref{gen_ghz}). Here an important point that must be noted is that for the parameter value $\theta$ $\in[0.43,1.13]$, the generalised GHZ state is found to be genuinely entangled.  
\\
\\
Now we consider tripartite W state which is a genuine entangled state, 
\begin{equation}
    \ket{W}=\frac{1}{\sqrt 3}(\ket{001}+ \ket{010} + \ket{100}).
\end{equation}
 To detect W state, we introduce the generalised map $\tilde{\Phi}_{3}$, where 
 \begin{equation}	 \label{modified}
 \begin{split}
      \tilde{\Phi}_{3}(*) & = [\tilde{\sigma_x}\circ\mathbb{P}_{1} \otimes \mathbb{I}_{2} \otimes \mathbb{I}_{3}+ \mathbb{I}_{1} \otimes \tilde{\sigma_x}\circ\mathbb{P}_{2}  \otimes \mathbb{I}_{3} \\
      & + \mathbb{I}_{1} \otimes \mathbb{I}_{2} \otimes \tilde{\sigma_x}\circ\mathbb{P}_3 + \kappa_3 . I.Tr](*)
       \end{split}
	 \end{equation} 
  The map $\tilde{\sigma}_x\circ\mathbb{P}$ denotes projection followed by a unitary operation $\sigma_x$. As local quantum operations such as unitary application cannot increase entanglement, therefore $\tilde{\Phi}_{3}$ gives positive output on all biseparable states. However, we will immediately discover that the modified version of the \emph{projection} map given by Eq.(\ref{modified}) would also be a GME map as $\tilde{\Phi_{3}}(\ket{W}\bra{W})$ gives us a negative eigenvalue -0.074 which is significantly smaller. However,  our proposed \emph{projection} map can detect both SLOCC classes of genuine tripartite entangled states. 

  Furthermore, we consider noisy W state as follows 
\begin{equation}
\widetilde{\rho_{noisy}}=x|W\rangle\langle W|+\frac{1-x}{8}{I}_8.
\end{equation} 	

To detect noisy W state with \emph{projection} map we apply the map as given in Eq.(\ref{modified}). Using the value of $\kappa_3$, the threshold value of $x$ for certifying GME of the given state turns out to be $x > 0.93$. \\

\textbf{Quadripartite system:}
Here. we take quadripartite GHZ state as:
\begin{equation}
\ket{GHZ}=\frac{1}{\sqrt 2}(\ket{0000}+\ket{1111}).
\end{equation} 
By computing the corresponding $\Phi_{4} (\rho)$ using the map defined in Eq.\eqref{quadripartitemap} where $\rho=|GHZ\rangle\langle GHZ|$ we obtain negative eigenvalue $-0.625$. Thus, \emph{ projection} map is capable of detecting genuine quadripartite entanglement. 

We also consider the robustness of the \emph{projection} map against noise by taking a 4-qubit noisy GHZ state 
\begin{equation}
\rho_{4-noisy}=x|GHZ\rangle\langle GHZ|+\frac{1-x}{16}{I}_{16}.
\end{equation} 	
The map $\Phi_{4}$ can detect $\rho_{4-noisy}$. It can be checked that the threshold value of $x$ for certifying GME of the given state turns out to be $x > 0.76$.
\subsection{Tripartite bound entangled state detection }
To achieve this goal, we consider a family of three-qubit bound entangled states in $\mathbb{C}^2\otimes \mathbb{C}^2 \otimes \mathbb{C}^2$ proposed in \cite{jafarizadeh2008detecting}. 

The state can be written as a linear combination of tensor products of identity and Pauli matrices \{$I_2, \sigma_x, \sigma_y, \sigma_z$\}. The state has the following expression:   
\begin{equation}
\begin{split}
    \tilde\rho  = & \frac{1}{8} \Big(I_{2} \otimes I_{2} \otimes I_{2} + 
 r_1 ({\sigma_z} \otimes {\sigma_z} \otimes I_{2}  + {\sigma_z} \otimes I_{2} \otimes   {\sigma_z}  \\ & +   I_{2} \otimes  {\sigma_z} \otimes   {\sigma_z}) +  r_2   {\sigma_x} \otimes  {\sigma_x} \otimes   {\sigma_x} +  r_3(  {\sigma_x} \otimes  {\sigma_y} \otimes   {\sigma_y} \\ & +   {\sigma_y} \otimes  {\sigma_x} \otimes   {\sigma_y} +  {\sigma_y} \otimes  {\sigma_y} \otimes   {\sigma_x})\Big)
 \end{split}
\end{equation}
where, $r_1 =  p_1 + p_2 - p_3$, $r_2= p_1 - p_2 + 3p_3 $,  $r_3=  -p_1 + p_2 +p_3$, and $0 \le p_{i} \le 1$ for all $i=1,2,3$. $p_1$, $p_2$ and $p_3$ satisfies a further constraint given by $p_1 + p_2 + 3p_3 =1$. To illustrate the efficacy of \emph{projection} map, we apply the map given by Eq.(\ref{tripartitemap}) on $\tilde\rho$ and find the eigenvalues of the output matrix. The eight eigenvalues turn out to be
\vspace{0.2cm}

$\lambda_1= \lambda_2= \lambda_3= \frac{1}{8} (7- p_1 -p_2 - 39p_3)$

$\lambda_4= \frac{1}{8} (7+ 43p_1 -37p_2 - 3p_3) $,

$\lambda_5= \frac{1}{8} (7- 37p_1 +43p_2 - 3p_3) $,

 $\lambda_6= \lambda_7= \lambda_8= \frac{1}{8} (7- p_1 -p_2 + 41p_3)$\\

 Now using the constraint $p_1 + p_2 + 3p_3 =1$, these eigen values reduce to 
 
 \vspace{0.2cm}
 $\lambda_1(p_1,p_2)= \lambda_2(p_1,p_2)= \lambda_3(p_1,p_2) =\frac{3}{4} (-1+ 2p_1+2p_2)$
 \vspace{0.05cm}
 
$\lambda_4(p_1,p_2)= \frac{1}{4} (3+ 22p_1 - 18p_2) $,
\vspace{0.05cm}

$\lambda_5(p_1,p_2)= \frac{1}{4} (3- 18p_1 + 22p_2) $, 
\vspace{0.05cm}

 $\lambda_6 (p_1,p_2)=\lambda_7 (p_1,p_2)=\lambda_8(p_1,p_2)= \frac{1}{12} (31 - 22p_1 - 22p_2)$
\vspace{0.05cm}
   \\
   
\begin{figure}[ht]
\includegraphics[width=.35\textwidth]{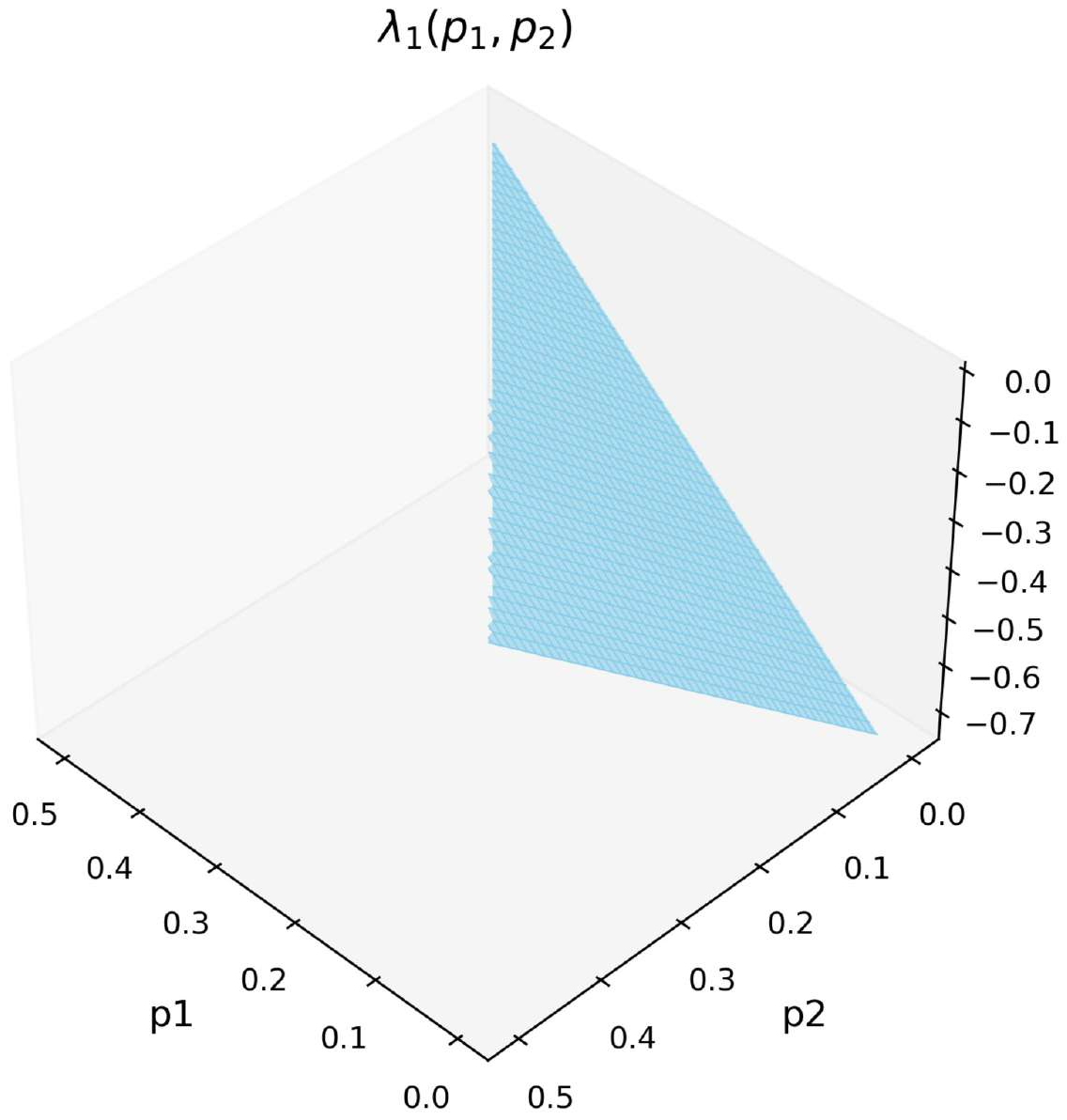}
\caption{ The negative region of the eigenvalue $\lambda_1(p_1,p_2)$ is depicted within the parameter space defined by $0 \le p_1 < \frac{1}{2}$ and $0 \le p_2 < \frac{1}{2} (1-2p_1)$.}\label{fig1}
\centering
\end{figure}

\begin{figure}[ht]
\includegraphics[width=.35\textwidth]{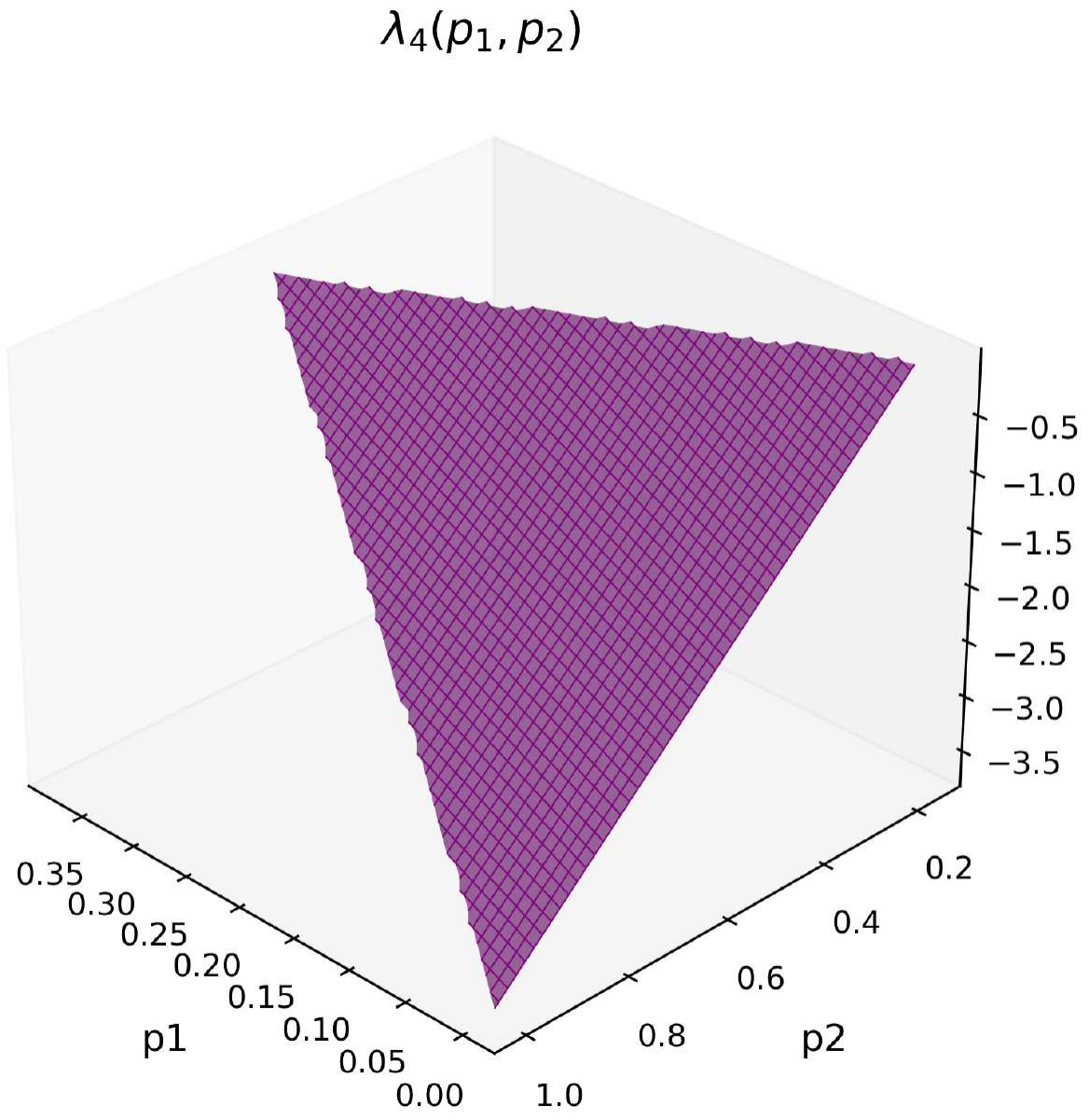}
\caption{The negative region of the eigenvalue $\lambda_4(p_1,p_2)$ is depicted within the parameter space defined by $0 \le p_1 < \frac{3}{8}$ and $\frac{1}{18} (3+22p_1) \le p_2 < 1-p_1$}\label{fig2}
\centering
\end{figure}

\begin{figure}[ht]
\includegraphics[width=.35\textwidth]{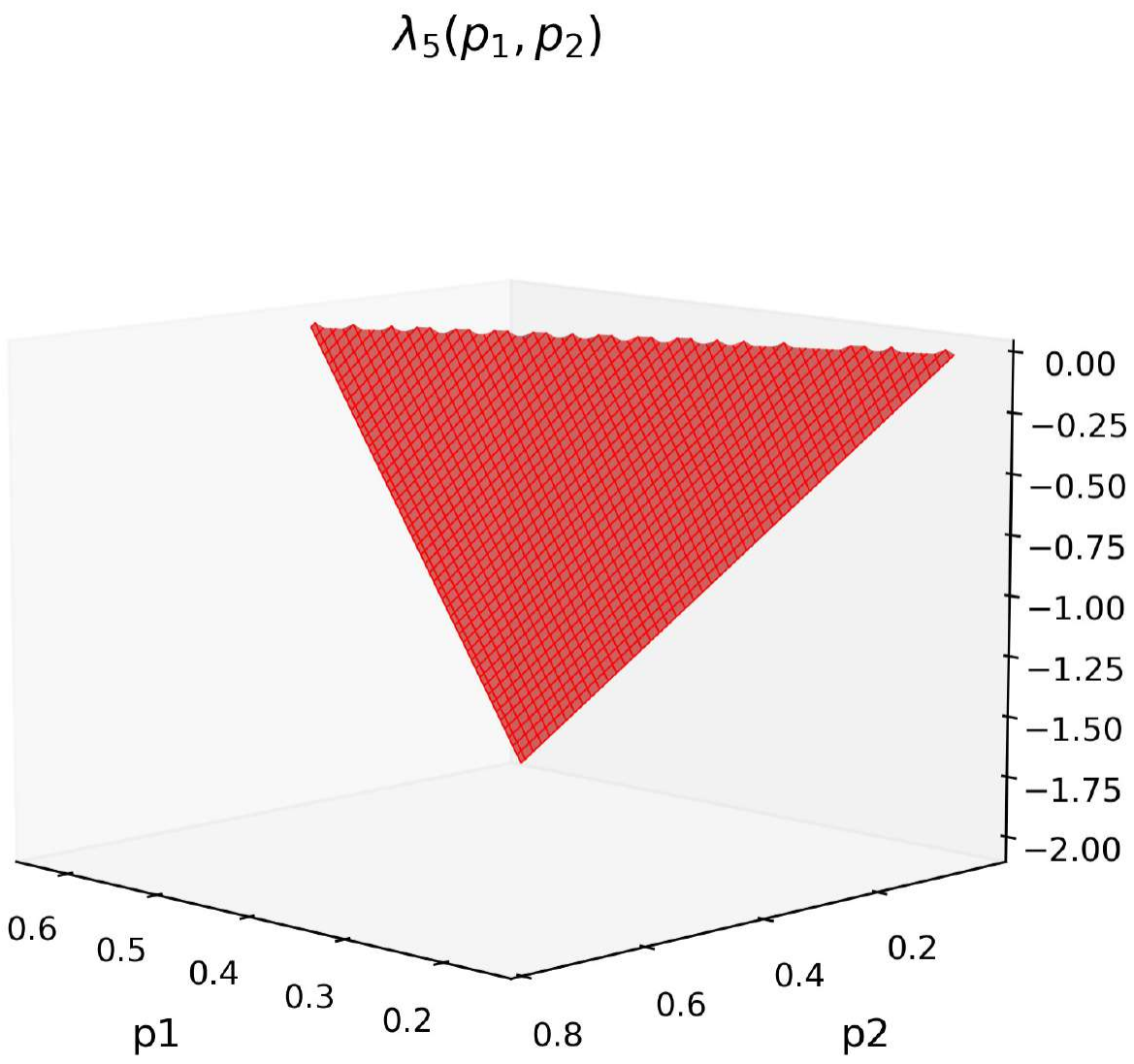}
\caption{The negative region of the eigenvalue $\lambda_5(p_1,p_2)$ is depicted within the parameter space defined by$\frac{1}{6} < p_1 \le \frac{5}{8}$ and $0 \le p_2 < \frac{1}{22} (-3+18p_1)$.}\label{fig3}
\centering
\end{figure}

It is to be noted that the eigenvalues $\lambda_6 , \hspace{0.2cm} \lambda_7 , \hspace{0.2cm} \lambda_8$ can never be negative since  $0 \le p_{i} \le 1$ for all $i=1,2,3$.
Therefore, the possible  negative eigenvalues are $\lambda_1(p_1,p_2), \hspace{0.2cm} \lambda_4(p_1,p_2), \hspace{0.2cm} \lambda_5(p_1,p_2)$.  
We plot these three eigenvalues in Fig. (\ref{fig1}), \hspace{0.1cm} Fig.(\ref{fig2}), \hspace{0.1cm} Fig. (\ref{fig3}) respectively which show its negativity in a particular region in the parameter space defined by $p_1$ and $p_2$. Consequently, we leverage an important aspect of \emph{projection} map that enables us to detect genuine tripartite entanglement as well as tripartite bound entangled states. Note that the partial transposition-based GME map \cite{clivaz2017genuine} is unable to detect this bound entangled state. Therefore, \emph{projection map} is capable of detecting a larger set of entangled states than the partial transposition-based GME map.

\vspace{5mm}

\vspace{5mm}
\vspace{5mm}

\subsection{Detection of a generic SLOCC class of quadripartite entangled state}

\begin{figure}[ht]
\includegraphics[width=.47\textwidth]{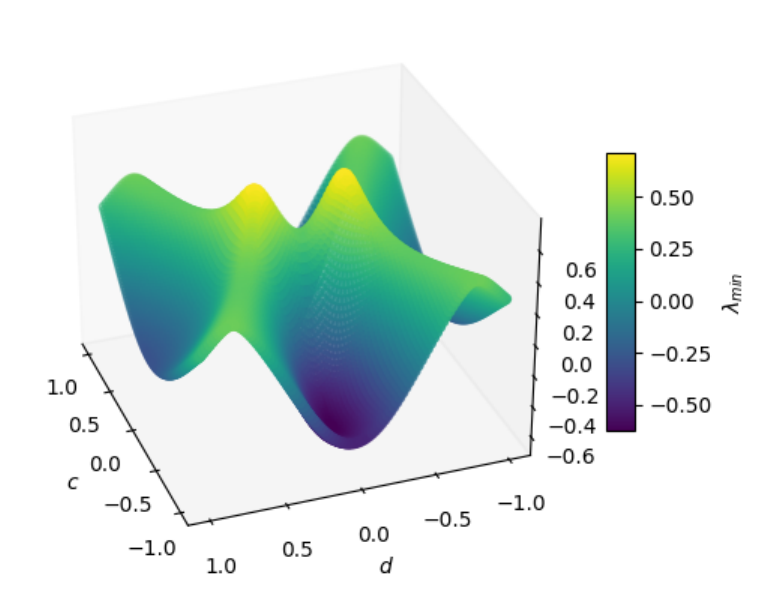}
\caption{ The negative region of the minimum eigenvalue $(\lambda_{min})$ of ${\Phi_{4}}(\ket{G_{abcd}} \bra{G_{abcd}})$ for the quadripartite state $\ket{G_{abcd}}$ is shown. The plot is obtained in the parameter region $a=d$ and $b=0.6$. }\label{gabcd}
\centering
\end{figure}

Next, we explore the potential usage of our framework for detecting genuine quadripartite entanglement by considering a SLOCC class of state  $\ket{G_{abcd}}$ introduced in \cite{verstraete2002four}
\begin{equation}
    \begin{split}
    \ket{G_{abcd}} & =\frac{a+d}{2}(\ket{0000}+\ket{1111})+\frac{a-d}{2}(\ket{0011}-\ket{1100}) \\ 
 + &\frac{b+c}{2}(\ket{0101}+\ket{1010})+\frac{b-c}{2}(\ket{0110}+\ket{1001})\\
 \end{split}
\end{equation}
where the coefficients $a$, $b$, $c$, and $d$ are complex parameters. The state is important due to the fact that a pure generic quadripartite state can be converted into $\ket{G_{abcd}}$. It is to be noted that for the parameter values given by $a=d=\frac{1}{\sqrt{2}}$, and $b=c=0$ the state reduces to GHZ state which is already shown to be detectable by our map. Now, we apply map $\Phi_{4}$ defined in Eq. \eqref{quadripartitemap} on the density matrix $\varpi=|{G_{abcd}\rangle\langle G_{abcd}|}$ and plot its minimum eigenvalues in Fig.(\ref{gabcd}) in the parameter region described by $a=d$, and $b=0.6$. It illustrates the region in which corresponding  ${\Phi_{4}}(\varpi)$ provides a negative eigenvalue. We emphasize our numerical method suggests that the positive partial transposition-based GME map and reduction-based GME map proposed in \cite{clivaz2017genuine} cannot detect $\ket{G_{abcd}}$. This reinforces the notion that the \emph{projection} map provides a more robust and comprehensive approach for detecting genuine entanglement in quadripartite systems.

\subsection{Genuine tripartite entanglement witness operator} Entanglement witness operators provide a practical and efficient method for detecting entanglement in quantum systems. Entanglement witnesses can be viewed as an application of the celebrated Hahn-Banach theorem in functional analysis \cite{holmes2012geometric}. This theorem states that any point lying outside of a convex and compact set can be separated by a hyperplane. A GME witness operator $(\mathcal{W})$ is a hermitian operator having at least one negative eigenvalue and satisfies $Tr(\mathcal{W} {\rho}_{2-sep}) \ge 0$  for all biseparable states ${\rho}_{2-sep}$ and $Tr(\mathcal{W} \sigma) < 0$ for at least one genuine entangled state $\sigma$.
\\
\\
\textbf{Proposition 2:} $\mathcal{W} = \Phi_{3}(\ket{GHZ}\bra{GHZ})$ acts as a witness corresponding to the map defined in Eq. \eqref{tripartitemap}.
 \begin{proof}
Let, $\rho_{2-sep}$ be a biseparable state defined in Eq. \eqref{bisep}. It can be checked that,
\begin{equation}
\begin{split}
   & Tr[\Phi_{3}(\ket{GHZ}\bra{GHZ})\hspace{0.1cm} \rho_{2-sep}] \\
   & = Tr[\Phi_{3}(\rho_{2-sep}) \ket{GHZ}\bra{GHZ}]
    \end{split}
\end{equation}
Now, from Eq.\eqref{positiveonbisep}, we know that 
\begin{equation}
    \Phi_{3}(\rho_{2-sep}) \ge 0
\end{equation}
Therefore, 
\begin{equation}
    Tr[\Phi_{3}(\rho_{2-sep}) \ket{GHZ}\bra{GHZ}] \ge 0
\end{equation}
which implies 
\begin{equation}
     Tr[\Phi_{3}(\ket{GHZ}\bra{GHZ})\hspace{0.1cm} \rho_{2-sep}] \ge 0
\end{equation}
Hence, $\mathcal{W}$ satisfies $Tr(\mathcal{W} \rho_{2-sep}) \ge 0$. Now, we want to show that $Tr(\mathcal{W} \sigma) < 0$ for at least one genuine entangled state $\sigma$.\\

Let, \begin{equation}
    \Tilde{\ket{GHZ}}= \frac{1}{\sqrt 2}(\ket{000}-\ket{111})
\end{equation}
Applying  $\mathcal{W}$ on $ \Tilde{\ket{GHZ}}$, we get 
\begin{equation}
    Tr(\mathcal{W} \hspace{0.1cm}  \Tilde{\ket{GHZ}}{ \Tilde{\bra{GHZ}})} = - \frac{1}{4}
\end{equation}
Therefore, the witness operator can detect $ \Tilde{\ket{GHZ}}$, which proves that $\mathcal{W}$ is a genuine tripartite entanglement witness.
\end{proof}
For experimental implementation of the witness, it is desirable to decompose it into a number of local observables \cite{guhne2003investigating}. We find the following decomposition of $\mathcal{W}$
into tensor products of the Pauli matrices \begin{eqnarray}
\mathcal{W} &=& \frac{1}{8}
\Big(7 I_2 \otimes I_2 \otimes I_2 +
\sigma_z\otimes \sigma_z\otimes I_2
+\sigma_z \otimes I_2 \otimes \sigma_z+ \nonumber\\&& I_2 
\otimes \sigma_z \otimes \sigma_z
+ 3(\sigma_x \otimes \sigma_x \otimes \sigma_x - \sigma_x \otimes \sigma_y \otimes \sigma_y
-\nonumber\\&&  \sigma_y \otimes \sigma_x \otimes \sigma_y
-\sigma_y \otimes \sigma_y \otimes \sigma_x)\Big)
\end{eqnarray}
This decomposition needs only measurements of the \emph{four} correlations given by $\sigma_x \otimes \sigma_x \otimes \sigma_x$, $\sigma_x \otimes \sigma_y \otimes \sigma_y$, $\sigma_y \otimes \sigma_x \otimes \sigma_y$, and $\sigma_y \otimes \sigma_y \otimes \sigma_x$. Measurements of the Pauli spin observables are experimentally feasible \cite{bourennane2004experimental}.  Note that one can similarly construct witness operator using \emph{projection} map for quadripartite and multipartite systems. Thus, the ingenuity of the \emph{projection} map provides a more simpler way for detecting GME rather than relying on quantum state tomography that might require $27$ measurements for a tripartite state. It is to be noted that the latter method of certifying entanglement of a general $N$-qubit state requires  $3^N$ measurements. In contrast, we have shown our method needs significantly lesser number of measurements.

\section{Conclusion}\label{conclusion}
 In quantum information theory, multipartite entanglement has been shown to be a resource for various information processing protocols \cite{cleve1999share,raussendorf2001one,dun2003security,durt2004security}. In this work, we have shown detection of the multipartite entanglement via \emph{projection} map which is a positive but not completely positive map.
 Our approach of certifying entanglement in multipartite scenarios has manifold implications. Firstly, we have constructed a suitable map by considering convex combination of the \emph{projection} map to detect genuine tripartite entanglement. 
Next, we formulate a general framework by taking convex combination of the \emph{projection} map
to detect GME of arbitrary $N$-qubit states. 
 Our present formalism based on the \emph{projection} map certifies both SLOCC classes of genuine tripartite entanglement \emph{i.e.} GHZ and W states. Moreover, our proposed map is shown to be capable of detecting tripartite bound entangled states. We have demonstrated this exclusive feature of \emph{projection} map by considering a family of states in $\mathbb{C}^2\otimes \mathbb{C}^2 \otimes \mathbb{C}^2$, and plotted the negative eigenvalue of the corresponding output matrix. It makes \emph{projection} map significantly advantageous than partial transposition-based GME map discussed in the context of multipartite entanglement detection \cite{clivaz2017genuine}.     
To illustrate the potential usage of our formalism, we consider a particular case of quadripartite state ($N=4$) and develop a suitable map to certify genuine quadripartite entanglement. 
We demonstrate that our proposed map can be used to detect a generic SLOCC class of quadripartite state. It makes \emph{projection} map quite distinctive in contrast to the earlier works devoted to detecting multipartite entanglement \cite{chimalgi2023detecting,clivaz2017genuine}. 
Secondly, our formalism leads to the construction of a suitable witness operator that can be implemented experimentally with a few measurement settings. It reveals that our construction provides a significant improvement over quantum state tomography based entanglement detection criterion that requires enormous measurements.

 As an interesting future direction, it would be interesting to construct a suitable measure of genuine multipartite entanglement that might involve \emph{projection} map operation. Also, it would be worthwhile to explore further usages of the \emph{projection} map for certifying higher dimensional entanglement, specifically bipartite and multipartite qudit systems. It is to be noted that one needs three Bloch components to specify a qubit state, whereas, the numbers rise to eight for $d=3$. But, \emph{projection} map has an advantage, since it reduces the corresponding Bloch components thereby minimizing the number of parameters in the state space. This suggests the feasibility of \emph{projection} map for further exploration towards detection and characterization of higher dimensional entanglement. 
\\
\section{Acknowledgement}
The authors acknowledge Prof. A.S. Majumdar for his valuable insights. BM acknowledges Bihalan Bhattacharya and Pritam Roy for fruitful discussions. BM acknowledges DST INSPIRE fellowship program for financial support.

\bibliography{Genuine}
	
\end{document}